\documentclass
[preprint,aps,showpacs,showkeys,preprintnumbers,amsmath,amssymb]
{revtex4}
\usepackage[dvips]{graphicx}

\def\be{\begin{equation}}
\def\ee{\end{equation}}
\def\bea{\begin{eqnarray}}
\def\eea{\end{eqnarray}}
\def\beas{\begin{eqnarray*}}
\def\eeas{\end{eqnarray*}}

\begin{document}

\title{Antibound poles in cutoff Woods--Saxon and
in Salamon--Vertse potentials}

\author  {J. Darai$^1$, A. R\'acz$^2$, P. Salamon$^{3}$, and R. G. Lovas$^{3}$}
\affiliation{$^1$Department of Experimental Physics, 
University of Debrecen,\\
PO Box 105, H--4010 Debrecen, Hungary\\
$^2$Faculty of Informatics, University of Debrecen, PO Box 12, 
H--4010 Debrecen, Hungary\\
$^3$Institute of Nuclear Research of the Hungarian Academy of Sciences,
PO Box 51, H--4001 Debrecen, Hungary\\
}
\date{\today}

\begin{abstract}
The motion of $l=0$ antibound poles of the $S$-matrix with varying 
potential strength is calculated in a cutoff Woods--Saxon (WS) potential  
and in the Salamon--Vertse (SV) potential, 
which goes to zero smoothly at a finite distance. 
The pole position of the antibound states as well as of the resonances 
depend on the cutoff radius, especially for higher node numbers.
The starting points 
(at potential zero) of the pole trajectories correlate well with 
the range of the potential. 
The normalized antibound radial wave functions on the imaginary 
$k$-axis below and above the coalescence
point have been found to be real and imaginary, respectively.   
\end{abstract}

\pacs{21.10.Pc,21.30.-x,21.60.Cs}

\keywords{$S$-matrix}

\maketitle

\section{Introduction}

Nuclear states are most often described in terms of single-particle (s.p.) 
bases generated by a spherical potential, mostly of Woods--Saxon (WS) type. 
Bound and discrete unbound s.p. states all obey the outgoing-wave 
boundary condition, which is $u(r,k)\sim\exp(ikr)$ when both the charge and 
the angular momentum $l$ are 0. The general solution behaves like 
$\exp(-ikr)-S\exp(ikr)$, 
where $S$, a function of the energy $E$ or the wave number $k$,  
is called the S-``matrix''. 
Where the outgoing boundary condition is satisfied, the S-matrix has poles. 
The bound-state poles belong to $E<0$
or imaginary wave number with ${\rm Im}\, k\equiv\gamma>0$. The resonance 
poles belong to complex $E$ and $k$, with 
$k=\pm\kappa-{\rm i}\gamma$ ($\kappa,\,\gamma>0$). For antibound (virtual) 
states, $E<0$, $k=-{\rm i}\gamma$ ($\gamma>0$). 

A WS basis is only complete if, in addition to bound states,
it contains continuum scattering states and/or resonances and/or 
antibound states
\cite{[Be68],[Mi09]}. The completeness is understood with respect 
to a generalized scalar product. 
The resonance states, which have definite intuitive meanings, 
have proved to be very useful in describing weakly bound or 
unbound states of nuclei \cite{[Mi09]}, unlike antibound states, 
whose exponential tail, $\exp(\gamma r)$, looks unphysical. However, 
the inclusion of an antibound state of $^{10}$Li \cite{[Ma00]} 
in the description of $^{11,12}$Li was found meaningful 
\cite{[Lov04],[Bet04],[Bet05],[Xu11]}. This shows that antibound states 
and the corresponding S-matrix poles (``antibound poles'') 
do deserve some attention. 

As an extension of recent studies \cite{[Sa08],[ra11]} of the dependence 
of the S-matrix poles on the tail behavior of the potential, we now study 
antibound poles. The nuclear potential should in principle have 
an exponentially decreasing tail, like the folding of the nuclear matter 
density with the one-pion exchange force. The standard WS potential 
obeys this criterion, but it can only be treated properly in analytical 
calculations, and analytical solution to the Schr\"odinger equation with 
a WS potential \cite{[Ben66]} only exists for angular momentum zero. 
The matter is that in numerically solving the problem with a prescribed 
boundary condition the solution has to be matched, at a finite distance, 
to the solution with potential zero (asymptotic solution), and the 
matching amounts to cutting off the tail of the potential. 
The error committed in this way is usually believed to be small, but 
in a recent paper it was shown that, for broad resonances, the poles in a 
cutoff WS potential strongly depend on the value of the cutoff radius
\cite{[Sa08],[ra11]}.  

In this work we examine the effect of the cutoff on the WS 
potential, and compare its behavior with a potential that goes to zero 
smoothly and is exactly zero beyond a point.
This ``Salamon--Vertse (SV) potential'' contains as many parameters 
as a cutoff WS potential and its shape is similar except for 
its tail. The tail of the SV potential can only conform to that of the WS 
at the expense of the inner region. Conformity in a longer section can 
be achieved with more parameters. This paper is only concerned with 
pointing out where problems might appear because of the cutoff. 

Unlike in former studies of antibound states we are aware of  
\cite{[Nu59],[Mi09]}, we now explore the wave functions as well. 
We limit our attention to $l=0$ since antibound states may only play 
some role for s-states.

\section{Potentials}
 
In solving the radial Schr\"odinger equation, a numerically 
calculated inner solution has to be matched at a distance 
$r=r_{a}$ to the solution of the asymptotic equation, 
and that yields the $S$-matrix. This procedure is tantamount to 
cutting off the WS potential at $r=R_{\rm max}\leq r_{a}$.
The potentials will be given in a form that expresses that they are exactly 
zero beyond a point, i.e., they are of finite range in a strict sense. 
The cutoff WS potential is thus 
\begin{equation}
\label{WSpot}
V^{\rm WS}(r)=V_0f^{\rm WS}(r),
\end{equation}
with
\begin{equation}
\label{vagottWS}
f^{\rm WS}(r)=\left\{
\begin{array}{rl}
-\left(1+{\rm e}^{\frac{r-R}{a}}\right)^{-1}
&\textrm{, if } r<R_{\rm max}\\
0~~~~&\textrm{, if } r\geq R_{\rm max}.
\end{array}
\right.
\end{equation}
In the resonance region the pole trajectories obtained by varying $V_0$ 
do depend on the cutoff radius $R_{\rm max}$ \cite{[ra11]}.  

The SV potential becomes zero  
beyond a finite value $r\ge\rho_{0}$ such 
that all its derivatives are also zero. Thus 
the potential is differentiable 
in the whole domain $r\in [0,\infty)$,
in contrast with the cutoff WS potential, which has a discontinuity at the
cut.

To follow Eq.~(\ref{WSpot}), we write the SV potential as 
\begin{equation}
\label{SVpot}
V^{SV}(r)=V_0 f^{SV}(r),
\end{equation}
where 
\begin{equation}
\label{newcent4}
f^{SV}(r)\equiv f^{SV}(r,c_1,\rho_0,\rho_1)=f_{\rho_0}(r) - c_1 
f^\prime_{\rho_1} (r), 
\end{equation}
with 
\begin{eqnarray}
\label{distrib}
f_{\rho}(r)&=&\left\{
\begin{array}{rl}
-{\rm e}^{\frac{r^2}{r^2-\rho^2}}   &\textrm{, if } r<\rho \\
0~~~~&\textrm{, if } r\geq\rho,
\end{array}
\right.
\\
\label{distribder}
f^\prime_{\rho}(r)&=&\left\{
\begin{array}{rl}
{\frac{2 r \rho ^2}{(r^2-\rho^2)^2}}{\rm e}^{\frac{r^2}{r^2-\rho^2}} &\textrm{, if } r<\rho \\
0~~~~~~~~&\textrm{, if } r\geq\rho.
\end{array}
\right.
\end{eqnarray}
The range parameters $\rho_0$ and $\rho_1$ are chosen as $\rho_0>\rho_1$, 
thus the potential in Eq.~(\ref{SVpot}) vanishes at $\rho_{0}$. To make 
the SV potential conform to the WS potential, we fit its three parameters, 
$\rho_0$, $\rho_1$ and $c_1$ ($c_1>0)$, to the 
WS form $f^{\rm WS}(r)$ \cite{[ra11]}. 

To have several antibound poles in the same potential, we choose the 
neutron potential to represent 
a heavy nucleus, $^{208}$Pb. The values 
$R=1.27\times 208^{1/3}$ fm $=7.525$ fm, $a=0.7$ fm
were adopted \cite{[Cu89]}, with $R_{\rm max}=15$ fm.
The SV parameters giving the best fit to the WS shape are:
$c_1=0.997$, $\rho_0=10.963$ fm and $\rho_1=8.328$ fm \cite{[ra11]}.

\section{Wave functions}

Let us sketch briefly how the pole solutions of the radial equation are 
calculated. For $l=0$ the radial equation is 
\begin{equation}
\label{radialeq}
\frac{d^2 u(r,k)}{dr^2} + [k^2-U(r)]u(r,k)=0\, ,
\end{equation}
where $U(r)=(2\mu/\hbar^2)V(r)$. We introduce an intermediate distance 
$R_{\rm im}$, where the internal (``left'') and external (``right'') solutions 
are to be matched. 
The left solution is defined in the interval $r\in [0,R_{\rm im}]$ such that 
\begin{equation}
\label{leftzero}
u_{\rm left}(0,k)=0, \quad\quad 
\left.\frac{du_{\rm left}(r,k)}{dr}\right|_{r=0}\equiv u^\prime(0,k)=1.
\end{equation}
The right solution is defined in the interval $r\in [R_{\rm im},r_{\rm a}]$, 
where $r_{\rm a}$ is in the asymptotic region
($r_{\rm a}\ge R_{\rm max}$ and $r_{\rm a}\ge\rho_0$), 
so that the solution satisfy the boundary condition 
\begin{equation}
\label{righbc}
u_{\rm right}(r_{\rm a},k)={\rm e}^{ikr_{\rm a}} \quad 
\left(u^{\prime}_{\rm right}(r_{\rm a},k)=ike^{ikr_{\rm a}}\right).
\end{equation}
We integrate Eq.~(\ref{radialeq}) numerically starting from the origin up 
to $R_{\rm im}$ and from $r_{\rm a}$ down to $R_{\rm im}$.
The eigenvalue or pole position is defined as the $k$ value for which the 
left and right logarithmic derivatives
\begin{equation}
\label{lgder}
L_{\rm left}(k)=\frac{u_{\rm left}^\prime(R_{\rm im},k)}
{u_{\rm left}(R_{\rm im},k)}, 
\,\,\,
L_{\rm right}(k)=\frac{u_{\rm right}^\prime(R_{\rm im},k)}
{u_{\rm right}(R_{\rm im},k)} 
\end{equation}
are equal:
\begin{equation}
\label{transc}
L_{\rm left}(k_j)-L_{\rm right}(k_j)=0,
\end{equation}
where, for antibound states, 
$k_j=-i\gamma_j$ ($\gamma_j>0$), with 
$j$ denoting the sequence number of the state. If we introduce
the matching factor 
$a_{\rm left}=u_{\rm right}(R_{\rm im},k_j)/u_{\rm left}(R_{\rm im},k_j)$ 
of the left solution, the eigensolution \begin{equation}
\label{nonnormsol}
v(r,k_j)=\left\{
\begin{array}{rl}
a_{\rm left} u_{\rm left}(r,k_j)
&\textrm{, if } r<R_{\rm im}\\
u_{\rm right}(r,k_j)
&\textrm{, if } r\geq R_{\rm im} 
\end{array}
\right.
\end{equation}
obtained in this way is well matched but not normalized. 

Since $\int_{r_{\rm a}}^{\infty}\exp(2ikr)dr=\infty$ for $k=-i\gamma$ 
with $\gamma>0$, the norm of an antibound state is infinity in the normal 
sense. By truncating the norm integral 
$\int_0^{\infty}v^2(r,k_j)dr$ at $r=r_{\rm a}$, 
the result will depend on $r_{\rm a}$ through a term 
$(2ik_j)^{-1}\exp(2ik_jr_{\rm a})$, 
which has to be eliminated. For resonance states 
this term can be eliminated either by using the prescription of 
Hokkyo~\cite{[Ho65]}, or by rotating the integration 
path of $\int_{r_{\rm a}}^{\infty}\exp(2ikr)dr$
onto the complex $r$-plane to the extent that the primitive function 
go to zero in infinity \cite{[GyV71]}, which results in 
$-(2ik)^{-1}\exp(2ikr_{\rm a})$ 
for the integral, and cancels the spurious dependence on $r_{\rm a}$ resulting 
from $\int_0^{r_{\rm a}}v^2(r,k_j)dr$. 
This rotation of the integration path 
provides a sound generalization for the scalar product involving
Gamow resonances \cite{[Be68]}, and makes it possible to construct 
complete sets involving resonance states. 
The same prescription also sets the tail term of the norm integral 
of an antibound state to $-(2ik)^{-1}\exp(2ikr_{\rm a})$ 
if a more radical rotation (by an angle $>\pi$) is applied, and the 
results with this formula are meaningful \cite{[Ve87]}. 
It is this 
prescription that allows the inclusion of antibound states in complete 
sets of states \cite{[Ve89]}. 
 
With this, the square of the norm of $v(r,k_j)$ is
\begin{equation}
\label{norm}
N^2= \int_0^{r_{a}}v^2(r,k_j) dr -C(r_{a},k_j),
\end{equation}
where 
\begin{equation}
\label{term}
C(r,k)=\frac{e^{2ikr}}{2ik }.
\end{equation}
The antibound wave function normalized to $1$ is thus 
\begin{equation}
\label{normwf}
u(r,k_j)=\frac{1}{N}v(r,k_j).
\end{equation}
For $k_j=-i\gamma_j$ ($\gamma_j>0$), the term $C(r_{a},k_j)$ is positive, 
just as the first term in Eq.~(\ref{norm}). Thus $N^2$ may be 
either positive or negative, {\em a fortiori} $N$ as well as $u(r,k_j)$ 
may be real or imaginary. Since the radial wave function $u$ enters 
in the norm integral as $u^2$, and $u$ must not be complex conjugated 
in any matrix elements \cite{[Be68]}, the imaginary wave function 
causes a strange behavior \cite{[Lov04]}. 

The pole positions $k_j$ and the corresponding normalized radial wave functions 
were calculated by a modified
version of the computer code GAMOW \cite{[Ve82]}.
The accuracy of the calculation was  checked by a more accurate program 
ANTI \cite{[Ix95]} using 
Ixaru's CP method \cite{[Ix84]}.

\section{Numerical results}

\subsection{Qualitative behavior of antibound poles}

\begin{figure}[t!]
\includegraphics[width=7cm]{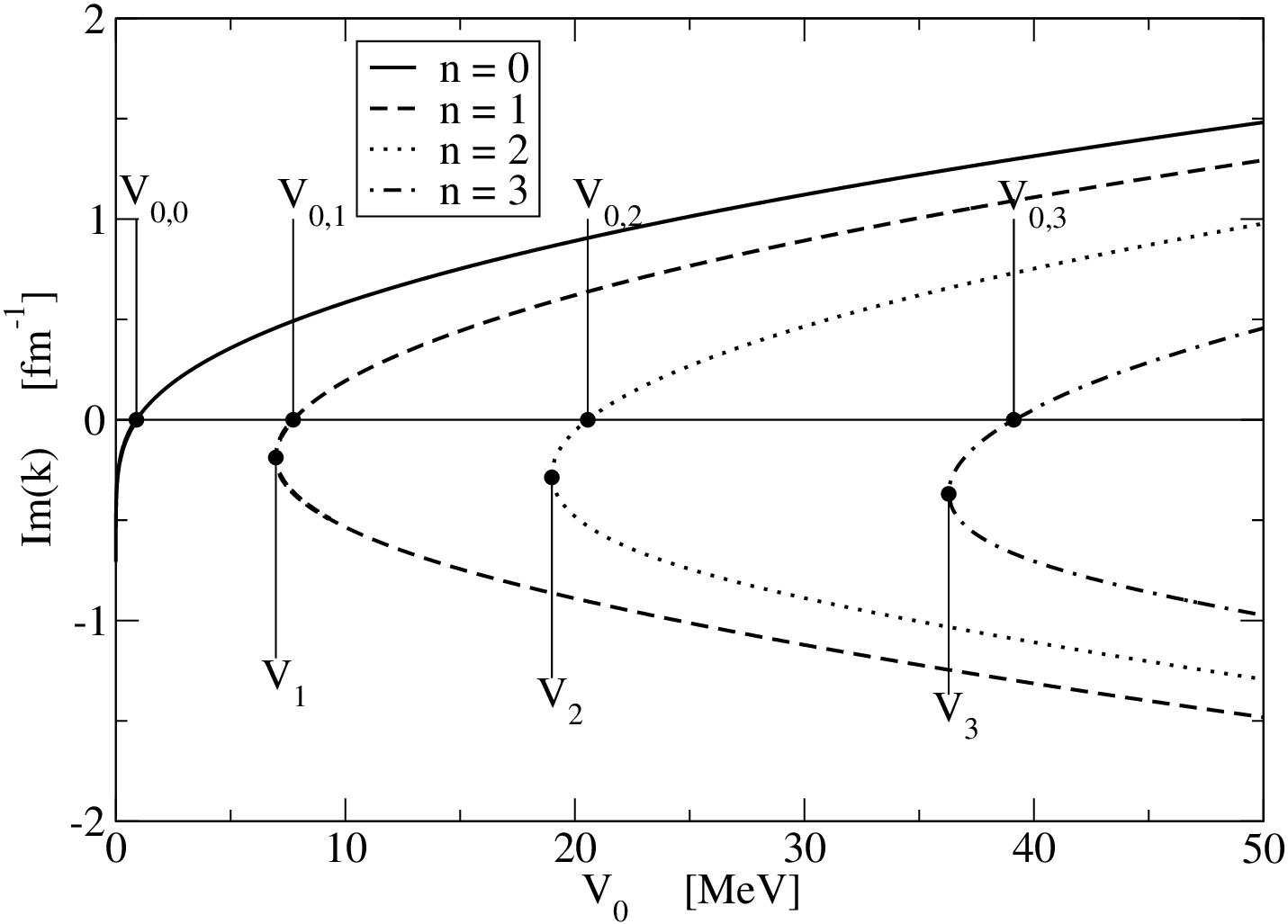}
\caption{Imaginary part of the pole wave number as a function of the depth of the WS potential. For bound and unbound states ${\rm Im}\,(k)>0$ and 
${\rm Im}\,(k)<0$, respectively.\label{abraws}}
\vspace*{1cm}
\includegraphics[width=7cm]{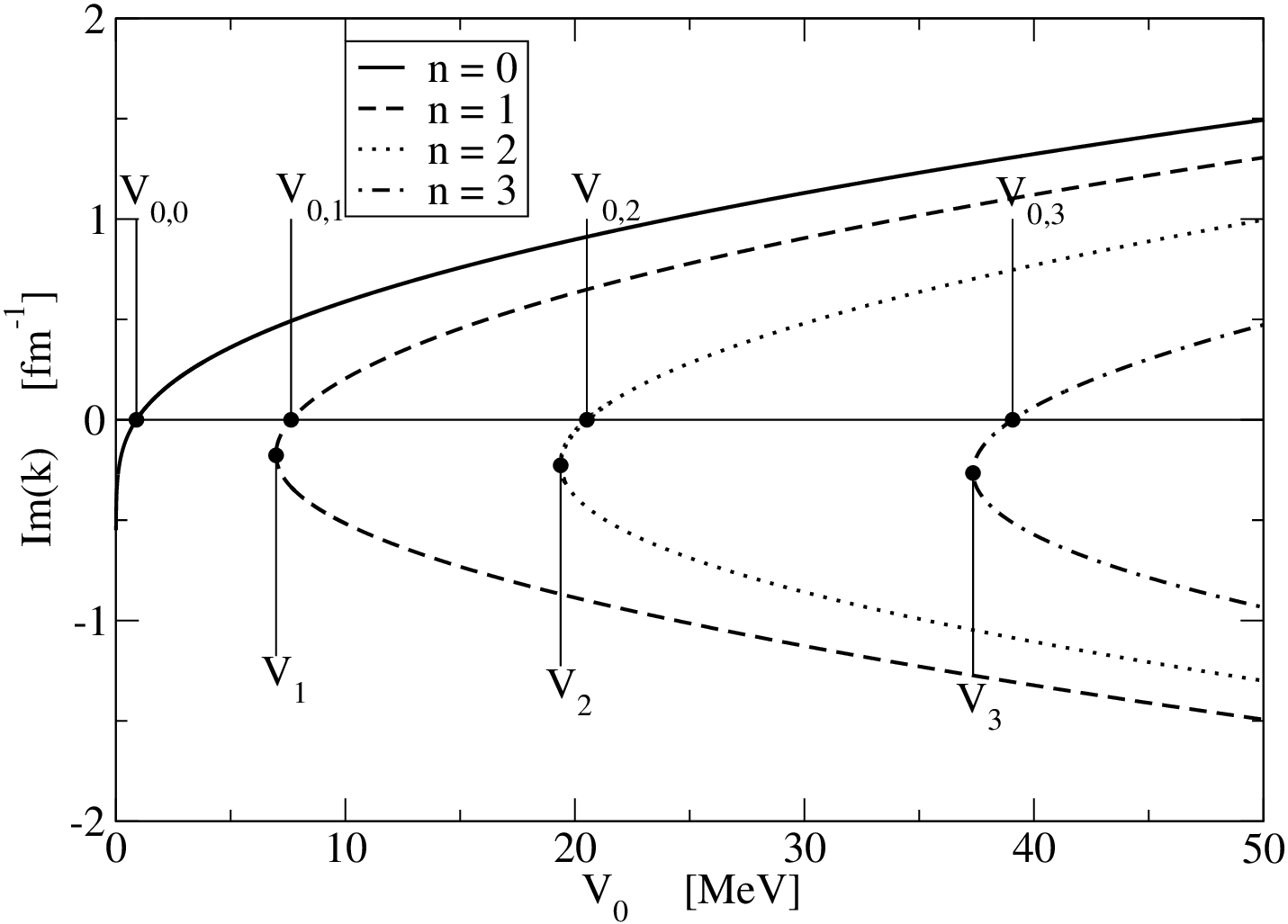}
\caption{Imaginary part of the pole wave number 
as a function of the depth of the SV
potential.\label{abrasv}}
\end{figure}

\begin{figure}[th]
\includegraphics[width=7cm]{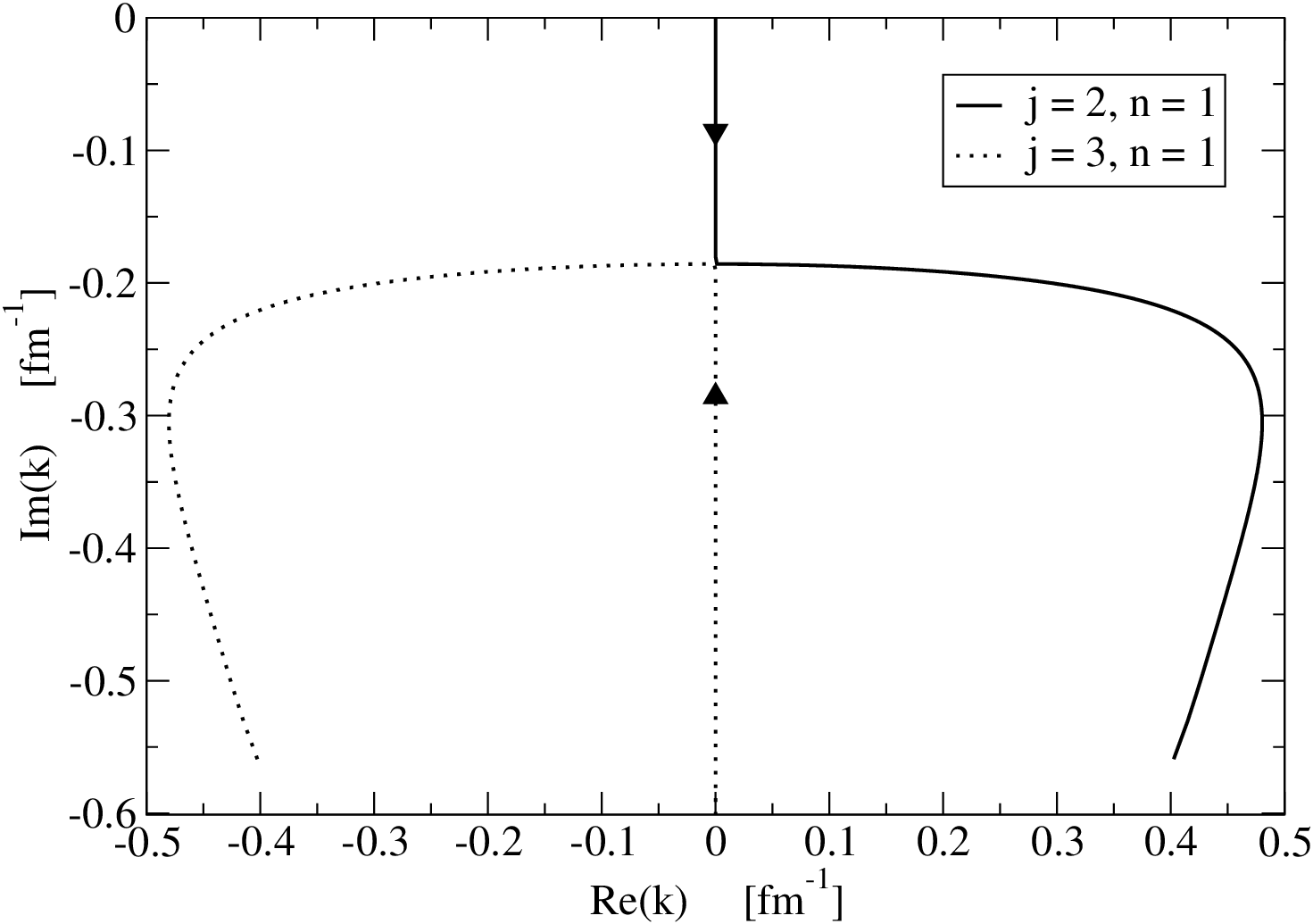}
\caption{Trajectories of the two $n=1$, $l=0$ poles in the WS
potential with $V_0$ varied ($R_{\rm max}=15$ fm).}
\label{polesmeet}
\end{figure}

Figures~\ref{abraws} and \ref{abrasv} show the imaginary part of the 
pole wave number $k$ as a function of the potential depth for the 
WS and for the SV potential, respectively. For bound and antibound 
states ${\rm Re}\,(k)=0$.
For a very shallow potential, there is just one antibound state, 
with node number $n=0$. With the attraction increased, 
the pole passes through the origin at $V_0=V_{0,0}$, and the system 
becomes bound. The ${\rm Im}\,(k)$ versus $V_0$ curves belonging to the 
other poles look like parabolas with horizontal axes. The bound states  
become antibound as the potential depth is decreased to $V_{0,n}$, 
and meet another antibound pole at $V_n$. What happens beyond  
their coalescence can only be depicted on the complex $k$-plane 
(Fig.~\ref{polesmeet}). The two poles part the ${\rm Im}\,(k)$-axis 
perpendicularly in opposite directions \cite{[Ne82]}.

We thus see that, while the bound state poles all move upwards 
along the imaginary $k$-axis when 
the potential is deepened, some antibound states behave conversely. 
The energy shift caused by a perturbation $\delta V_0f(r)$ can be 
estimated by  $\delta E=\int_0^{r_{\rm a}} u^2(r,k_j)\delta V_0 f(r) dr$. 
The sign of $\delta E$ with respect to that of $\delta V_0f(r)$ 
depends on whether $u(r,k_j)$ is real or imaginary. 
By looking at the $V_0$ dependence of the pole, we can 
unambiguously infer that the wave function is imaginary on the 
upper branches of the parabolas, and they are real below. The single 
$n=0$ antibound-state wave function is imaginary.

\begin{figure}[hb]
\vspace*{0.5cm}
\includegraphics[width=7cm]{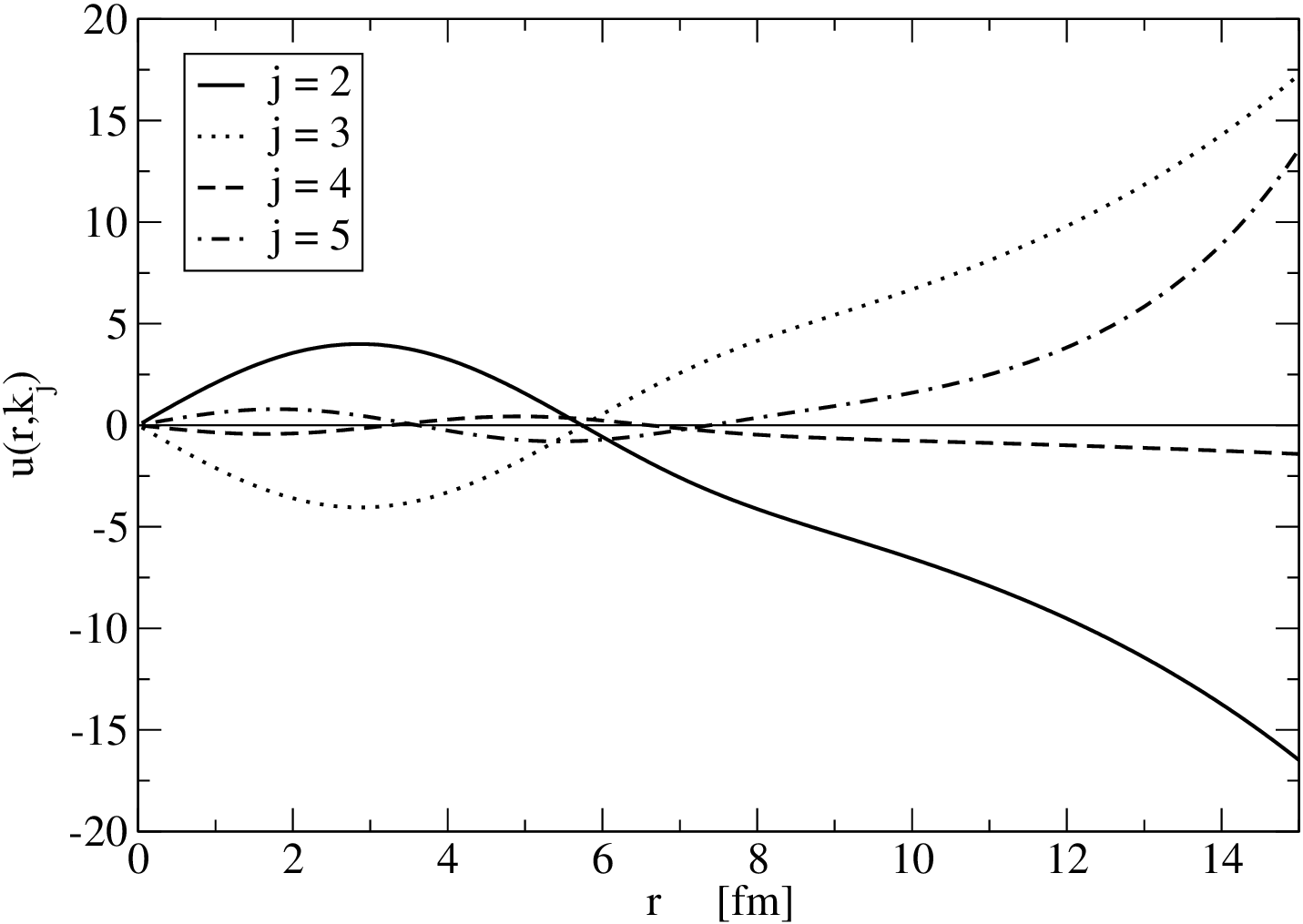}
\caption{Normalized radial wave functions of antibound states in WS 
potentials. The $n=1$ ($j=2,3$) and $n=2$ ($j=4,5$) states were produced 
by $V_0=6.9692$ and 19.5 MeV, respectively. 
The wave numbers $k_j$ (in fm$^{-1}$) are $k_2=-0.183\,{\rm i}$, 
$k_3=-0.188\,{\rm i}$, $k_4=-0.410\,{\rm i}$, $k_5=-0.422\,{\rm i}$. 
The functions with $j=2,4$ are imaginary, while those with $j=3,5$ 
are real.}
\label{radwfn}
\end{figure}

In Fig.~\ref{radwfn} we show the radial wave functions of some normalized 
antibound states in WS potentials. 
The antibound states that belong to the same node number 
in two different branches of the parabola seem to be non-orthogonal 
to each other although they are generated, pairwise, by the same potential. 
That is, however, just an appearance. In fact, the tail region of the 
overlap integral cancels the the contribution of the inner region. 
The $j=2$ and $j=3$ antibound states are orthogonal to each other, and so are 
the $j=4$ and $j=5$ states. Thus, pairwise, they may be included 
in complete sets of states \cite{[Be68]} simultaneously. (The 
antibound states of different node numbers are, of course, orthogonal 
to each other if, unlike in Fig.~\ref{radwfn}, they are produced 
by the same potential.)

If we have a centrifugal or Coulomb barrier, the picture is 
different in that the bound-state poles meet the antibound poles at the origin, 
and bifurcate there into a pair of resonance poles. In the 
($V_0$,${\rm Im}\,(k)$) plane this corresponds to 
parabolas whose apices are at the origin. When the potential bottom is lifted, 
the antibound poles approach the origin monotonously from below, thus 
their normalized wave function is real throughout.

\subsection{Quantitative observations}

\begin{figure}[h]
\vspace{0.5cm}
\includegraphics[width=7cm]{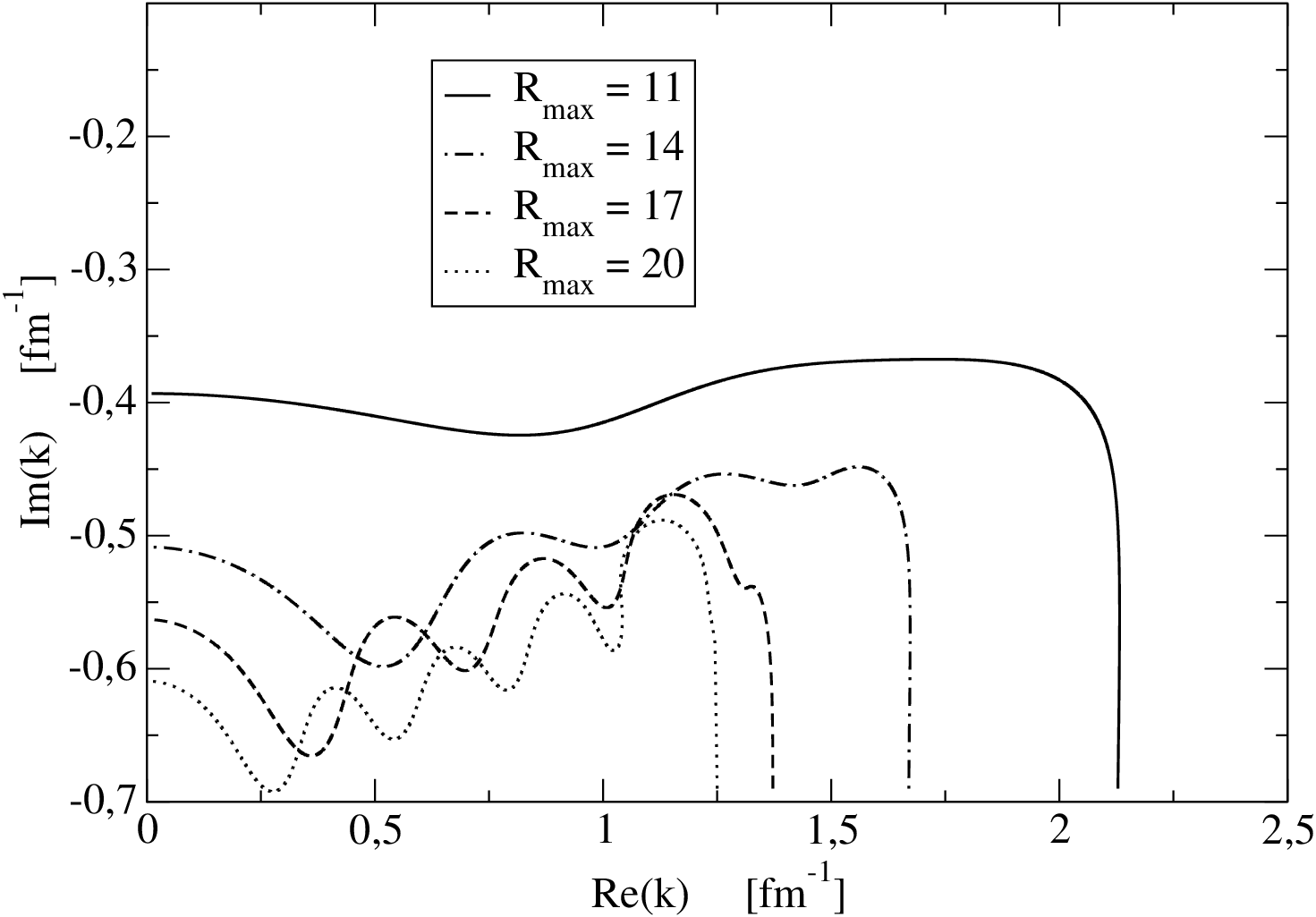}
\caption{ $n=7$ resonance trajectories 
for the WS potential cut off at different $R_{\rm max}$ values. }
\label{wsn6trajs}
\end{figure}

A numerically most sensitive quantity is the apex $V_n$ of the parabolas
in Fig.~\ref{abraws}, and that was used for testing 
the $R_{\rm max}$-dependence for the WS potential (Table~\ref{WSdepths}).    
We see that for $n=1,2,3$ the $V_n$ values are practically independent 
of $R_{\rm max}$. The largest variation is in $V_3$, most probably 
due to the enhancement of the error in the numerical solution 
of the differential equation as discussed in Ref.~\cite{[Bo95]}.
The $k$-values of the apices are somewhat more sensitive to 
$R_{\rm max}$, and the sensitivity gets more pronounced for higher $n$. 
(We will return to this problem in discussing $R_{\rm max}$-dependence of 
the pole trajectories, see Fig.~\ref{wsn6trajs} later.) 
The sensitivity to the potential shape has also been tested by comparing the 
values obtained for the potential strength $V_{0,n}$, which puts the pole 
at the threshold (Table~\ref{crossdepths}). For WS, $R_{\rm max}=15$ fm 
was used, but it was ascertained that $V_{0,n}$ is practically independent 
of $R_{\rm max}\in [15,25]$ fm. The strengths for the two potential forms 
are very similar, which follows from the shapes being very similar.

\begin{table}[ht]
\begin{center}
\caption{Well depths $V_n$ at the coalescence of the two 
antibound poles 
\label{WSdepths}}\begin{tabular}{lcccc}
\hline
&$R_{\rm max}$ (fm) &$V_1$ (MeV) &$V_2$ (MeV) &$V_3$ (MeV)\\
\hline
WS&15&6.969&18.995&36.286\\
  &16&6.969&18.992&36.263\\
  &18&6.968&18.989&36.239\\
  &20&6.968&18.989&36.239\\
  &25&6.968&18.989&36.230\\
\hline
SV&&6.978&19.378&37.347\\
\hline
\end{tabular}
\end{center}
\end{table}

\begin{table}[bh]
\begin{center}
\caption{The values $V_{0,n}$ setting the pole at the threshold 
\label{crossdepths}}
\begin{tabular}{lcccc}
\hline
Potential&$V_{0,0}$ (MeV) &$V_{0,1}$ (MeV) &$V_{0,2}$ (MeV) &$V_{0,3}$ (MeV) \\
\hline
WS&0.897&7.727&20.562&39.122\\
SV&0.893&7.634&20.519&39.072\\
\hline
\end{tabular}
\end{center}
\end{table}

\begin{figure}[th]
\vspace{0.5cm}
\includegraphics[width=7cm]{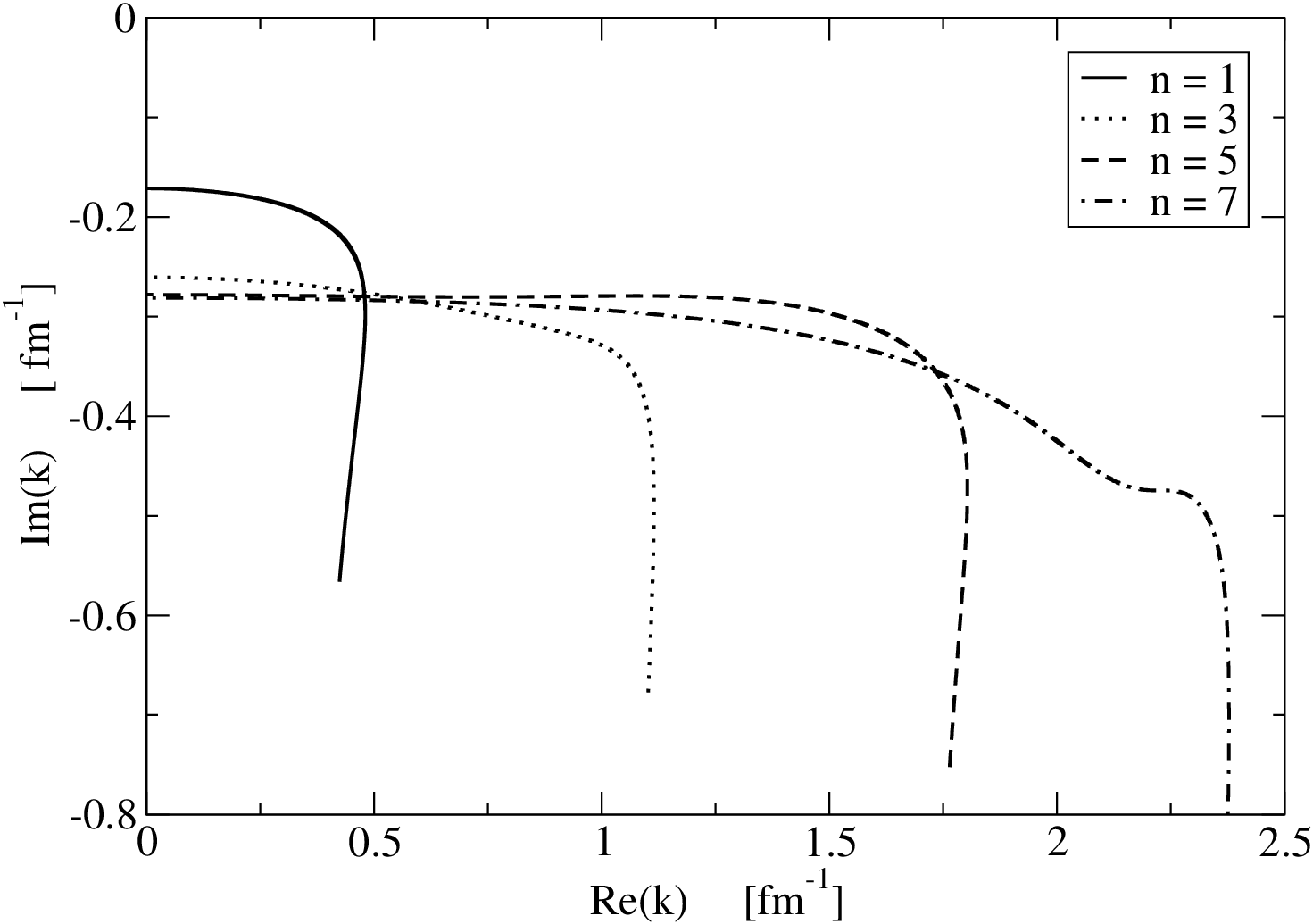}
\caption{Trajectories of the odd-$n$, $l=0$ poles in the SV 
potential with $V_0$ varied. The even-$n$ trajectories are similar, except for $n=0$, which runs along the imaginary $k$-axis.}
\label{l0svtraj}
\end{figure}

As we showed in Fig.~\ref{polesmeet}, beyond the coalescence, the pair 
of antibound poles is transformed into a pair of decaying and capturing 
resonance poles. We show the trajectories 
of some of the $l=0$ decaying resonances in the complex $k$-plane in 
Fig.~\ref{l0svtraj}. (The poles of the capturing resonances are the mirror 
images of the decaying ones with respect to the ${\rm Im}\,(k)$-axis.)

The starting point of a trajectory is defined by 
the limit $k_j=\lim_{ V_0 \to 0}\,{k_j(V_0)}$. For a potential of range 
$R$, an estimate 
for this limit is given by \cite{[Ne82]} 
\begin{equation}
\label{rekn}
{\rm Re}\,(k_{n})= \frac{n\pi}{R} + O(1).
\end{equation}
For large $n$ we can perhaps neglect the term $O(1)$. We 
approximate $V_0=0$ by 5 keV. 

In Fig.~\ref{l0svtraj} one can see the 
trajectories of the $l=0$, $n=1,3,5,7$ poles in the SV potential. 
Only the (anti)bound states have definite node numbers $n$, 
but resonances can also be characterized by the node number of the 
(anti)bound state that they correspond to. 
The real parts of the starting points are seen to be almost equidistant. 
Therefore, these ${\rm Re}\,(k_n)$ values can be fitted well
by the straight line: ${\rm Re}\,(k_n)=a_0+a_1n$, with a slope 
$a_1=0.32$ fm$^{-1}$, which implies $R=9.778$ fm.

\begin{figure}[hb]
\vspace{0.5cm}
\includegraphics[width=7cm]{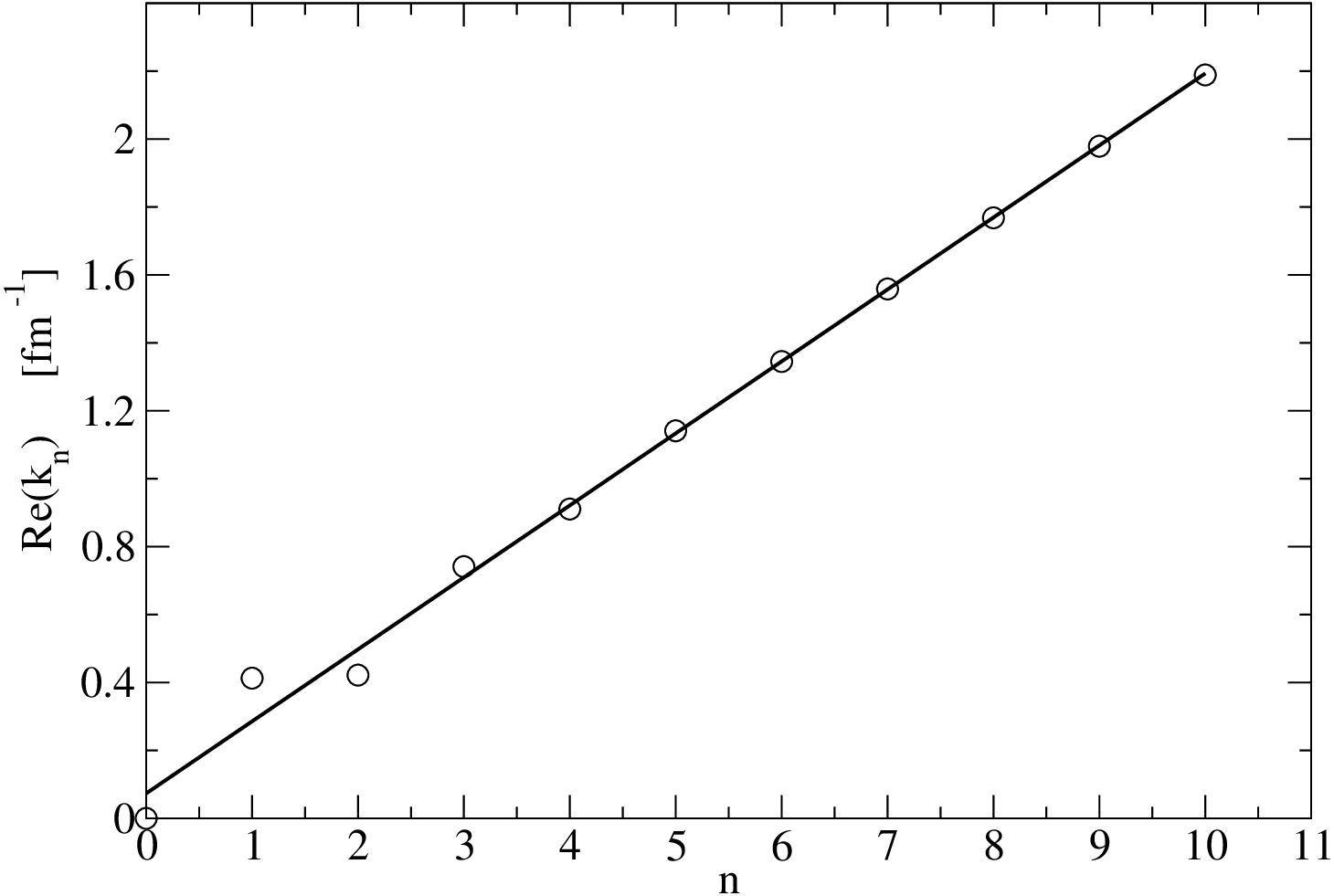}
\caption{${\rm Re}\,(k_n)$ values of starting points of
resonance trajectories 
for the WS potential cut off at $R_{\rm max}=15$ fm fitted with a  
straight line}
\label{reknn}
\end{figure}

As for the WS potential, in Fig.~\ref{reknn} one can see the starting points
${\rm Re}\,(k_n)$ as a function of $n$, and a straight line fitted to it. 
Although, for $n<4$, the ${\rm Re}\,(k_n)$ values are somewhat erratic, 
the slope of the line, $a_1=0.212$ fm$^{-1}$, provides $R=14.82$ fm,
in good agreement with $R_{\rm max}=15$ fm. To see the dependence of the 
$R$ value deduced in this manner on $R_{\rm max}$, we repeated the 
calculations for a set of $R_{\rm max}$ values chosen from the range typically 
used in practical calculations. The slope of the line was determined from 
five points with $n=4,\ldots,8$. The results are given in Table~\ref{derivR}.
The smaller $R_{\rm max}$, the better is the agreement with $R$, and the better 
is Eq.~(\ref{rekn}) satisfied. For larger $R_{\rm max}$ the rund-off errors 
of the numerical solution of the radial equation get larger.
This fact forbids one to go substantially beyond $R_{\rm max}=20$ fm. 

\begin{table}[ht]
\begin{center}
\caption{Ranges $R$ obtained from Eq.~(\ref{rekn}) for different values 
of the cutoff radii $R_{\rm max}$. The trajectory starting points 
${\rm Re}\,(k_n)$ were fitted by a linear function of $n$, and $R$ 
was calculated from its slope. The $\sigma$ values show the quality 
of the fit of the data to the straigt line. \label{derivR}}
\begin{tabular}{ccccc}
\hline
$R_{\rm max}$ (fm) &&$R$ (fm) &&$\sigma$\\
\hline
  11               &&10.93    &&$1.6\times 10^{-6}$\\
  14               &&13.85    &&$2.0\times 10^{-5}$\\
  17               &&16.89    &&$2.0\times 10^{-5}$\\
  20               &&20.45    &&$6.4\times 10^{-4}$\\
\hline
\end{tabular}
\end{center}
\end{table}

We examined the sensitivity of the pole trajectories to the cutoff radius, and 
in  Fig.~\ref{wsn6trajs} we illustrate the results with the case of $n=7$, 
which, for $R_{\rm max}=15$ fm, fits well into the straight line in 
Fig.~\ref{reknn}. In view of the approximate independence of $V_n$ on  
$R_{\rm max}$ (see Table~\ref{WSdepths}), the results look surprising. 
We see that the $n=7$ trajectory and, indeed, its point of intersection 
with the imaginary $k$-axis, depend appreciably on the cutoff.
While $R_{\rm max}$ is changed between 11 fm and 20 fm, the intersection of 
the trajectories with the Im$\,(k)$-axis moves from $-{\rm i}0.40$ fm$^{-1}$ 
to $-{\rm i}0.61$ fm$^{-1}$, with the potential depth to be set to 
166 and 159 MeV, respectively. Thus, similarly to the $n=1,\,2,\,3$ 
cases, $V_7$ is less sensitive to $R_{\rm max}$ than the pole positions.

The stability of $V_n$ as a function of $R_{\rm max}$ can be understood,  
again, in a perturbative picture. The shift of a pole energy caused 
by changing the cutoff radius $R_{\rm max}$ from $R_1$ to $R_2$ can be 
estimated to be $\Delta E=\int_{R_1}^{R_2}V^{\rm WS}(r)u^2(k_j,r)dr$. 
Now, the tail of $V^{\rm WS}(r)$ is small, but, for a resonance,  
$u^2(k_j,r)\sim{\rm e}^{2{\rm i}kr}$ (for $r\in[R_1,R_2]$) is complex and 
may take large absolute values, and, correspondingly, the resonance poles 
may be shifted appreciably in the complex $E$-plane as well as in the 
$k$-plane. For antibound poles, however, the function $u^2(k_j,r)$ is 
real, so that the pole can only be shifted along the imaginary $k$-axis. 
In the perturbative approximation the coalescence point of a resonance 
trajectory is thus shifted into the coalescence 
point of the shifted trajectory with unchanged $V_n$, 
which suggests that $V_n$ need not be changed much when 
$R_{\rm max}$ is varied even in an accurate calculation. 

Looking at the trajectories in Fig.~\ref{wsn6trajs}, we see that 
the larger the value of $R_{\rm max}$, the farther from the origin 
do they intersect with the Im$\,(k)$-axis. Moreover, near vanishing 
potential, all trajectories start with a vertical section 
at a certain ${\rm Re}\,(k_7)=\kappa_7$. The larger the value of $R_{\rm max}$, 
the smaller is $\kappa_7$, and the inverse proportionality expressed by 
Eq.~(\ref{rekn}) is borne out. 

It is interesting to compare this behavior with the case of the square-well 
potential explored in  Ref.~\cite{[Ba80]}.
For such a potential with radius $R$, the value of $\beta_n(R)=\bar\gamma_n R$ 
(with $-$i$\bar\gamma_n$ denoting the coalescence point) 
was found to be equal to 1, at least for low $n$ values, 
independently both of $R$ and of the node number $n$ \cite{[Nu59]}. 
For a cutoff WS potential the
corresponding $\beta_n=R_{\rm max}\bar\gamma_n$ does depend on $R_{\rm max}$,
and, for the $n=7$ case shown in Fig.~\ref{wsn6trajs}, can be approximated by
a first order polynomial: $\beta_7(R_{\rm max})=-5.05+0.864R_{\rm max}$.

\section{Summary}

We can summarize the results as follows.

The strange behavior of the antibound basis state found in 
Ref.~\cite{[Lov04]} is explained by its normalized radial wave function 
being imaginary. Except for $n=0$, the poles occur pairwise, 
and there is a range of potential depths in which there are two 
antibound states of the same node number: one below, and the other 
above the coalescence point. It has been shown that the antibound 
states lying below the coalescence
points are real, while those above are imaginary. 
This seems to be a general property of antibound wave functions.  
The antibound states may be included in an orthonormal basis. 
Numerical examples show that even those that belong to the same node 
number are orthogonal to each other.   

The pole belonging to node number $n=0$ is an exception; it 
starts (with an infinitesimally small attractive potential)  
as an antibound state, and becomes bound when the potential is deepened, 
without ever passing into the resonance region.
The behavior of all other poles show similarity to the $l>0$ case 
\cite{[ra11]}: the real parts of the starting points of the 
resonance trajectories (near potential zero) 
are inversely proportional to the potential range. 
For the WS potential, this range is to be identified with 
the cutoff radius. 
For the WS potential the pole trajectories, including the positions 
of the antibound states, depend on the cutoff radius, 
and the higher the node number, the stronger the dependence is. 
Thus, without discrediting the use of the WS potential in representing 
the nucleus in bound-state or scattering problems, this paper 
cautions against its indiscriminate use to represent broad resonances 
or antibound states.

\section*{Acknowledgment}

The authors are grateful to Prof. T. Vertse for valuable discussions.
This work was supported by the OTKA Grant No. K72357 and 
by T\'AMOP project 4.2.1./B-09/1/KONV-2010-0007/IK/IT.
The latter is co-financed by the European Social Fund 
and the European Regional Development Fund.

\end{document}